\documentclass[aps,twocolumn]{revtex4}
\usepackage{graphicx}
\usepackage{amsmath}
\newcommand{\be}{\begin{equation}}
\newcommand{\ee}{\end{equation}}
\begin{document}

\title{Experimental studies of the internal Goos-H\"anchen shift
for self-collimated beams \\in two-dimensional microwave photonic crystals}

\author{Aaron Matthews}
\author{Yuri Kivshar}

\affiliation{
Nonlinear Physics Centre and Centre for Ultra-high bandwidth Devices
for Optical Systems (CUDOS), Research School of Physical Sciences
and Engineering, Australian National University, Canberra ACT 0200,
Australia}

\begin{abstract}
We study experimentally the Goos-H\"anchen effect observed at the reflection of a self-collimated beam from the surface of a two-dimensional photonic crystal and describe a method for controlling the beam reflection through surface engineering. The microwave photonic crystal, fabricated from alumina rods, allows control of the output position of a reflected beam undergoing an internal Goos-H\"anchen shift by changing the rod diameter at the reflection surface. The experimental data is in good agreement with the results of the finite-difference time-domain numerical calculations.
\end{abstract}

\maketitle

Periodic photonic structures such as photonic crystals allow the substantial modification of both the dispersion and diffraction of electromagnetic waves. Many unique dispersion-modifying phenomena have been predicted and observed for photonic crystals including negative refraction, superprism effects, and light self-collimation.  In particular, electromagnetic waves propagating in homogeneous dielectric media diffract, and in order to preserve the size of a narrow beam one needs to trap light by employing different types of waveguiding structures. In contrast, photonic crystals allow almost diffractionless propagation of light under certain conditions, this is known as light self-collimation. When a photonic crystal operates in the self-collimation regime, electromagnetic waves propagate without diffraction-induced divergence even in the absence of nonlinear effects. The self-collimation of light beams in photonic crystals has attracted special attention due to many promising applications in photonic-crystal based structures and circuits~(see, e.g., Refs.~\cite{ref1,ref2,ref3,ref4,ref5,ref6,ref7,ref8,ref9,ref10,ref11} and references therein).

\begin{figure}[htb]
\centerline{\includegraphics[width=8cm]{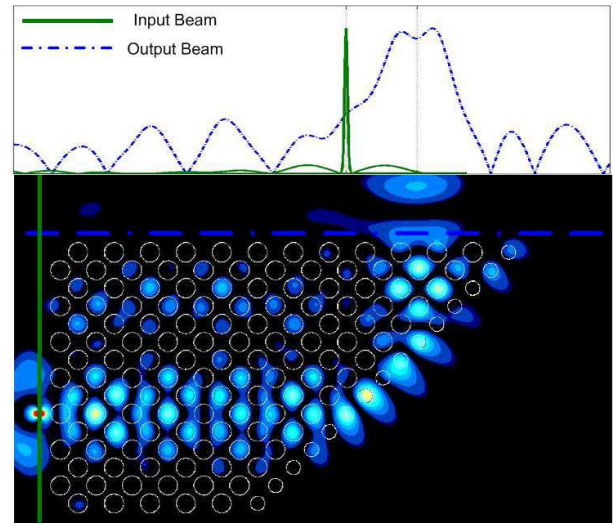}}
\caption{(Color online) Numerical FDTD simulations of a reflected self-collimated beam experienced a GH shift for
the modified surface of 5.92mm rods.  Top: input (green) and output (blue) normalized field profiles.
Bottom: Electric field generated by a point source and reflected by the surface.}
\end{figure}

The self-collimation effect originates from the strong anisotropic dispersion that makes the equal-frequency contours ultra-flat within certain angular and frequency ranges. Since the propagation of energy is always normal to the equal-frequency contours, the light only propagates along the direction normal to the equal-frequency contours inside the photonic crystal. Based on this self-collimation effect, some simple and effective bending and splitting concepts have been suggested~\cite{ref9,ref10,ref11}. In particular, it was shown that a surface of a truncated photonic crystal can operate as a total-internal-refection mirror for the self-collimated beams, and this effect can be employed to create sharp bends.

\begin{figure}[htb]
\centerline{\includegraphics[width=8cm]{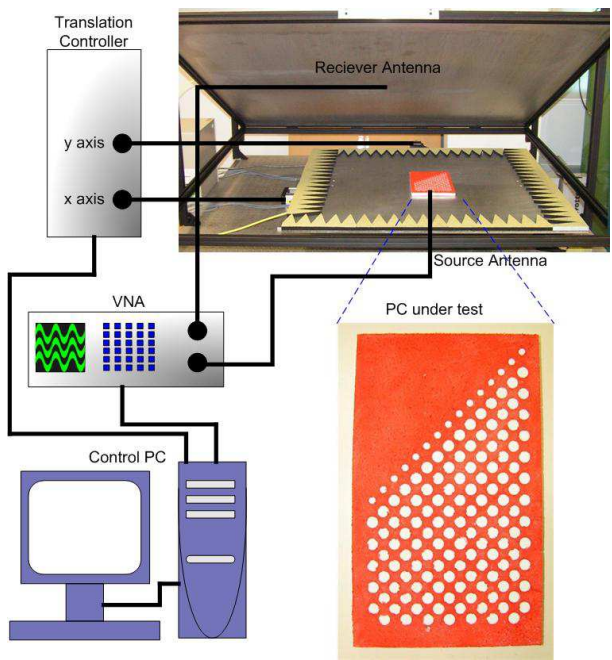}}
\caption{(Color online) Experimental setup. Included are photographs of the parallel-plate waveguide
mounted on a two-axis translation stage and the fabricated photonic crystal.}
\end{figure}

The surface of a truncated photonic crystal shows many interesting properties due to both the surface structure and its ability to support surface states. Surface states in photonic crystals have only been shown to exist under an appropriate change of the surface layer, such as a termination though the surface cell or a change in the surface geometry or material properties (see, e.g., Refs.~\cite{surf1,surf2,surf3}). In addition, the corrugation inherent to the surface of truncated photonic crystals leads to a number of interesting effects including beam shaping and enhanced beaming~\cite{beam1,beam2,Steven_beam}.

\begin{figure}[htb]
\centerline{\includegraphics[width=9cm]{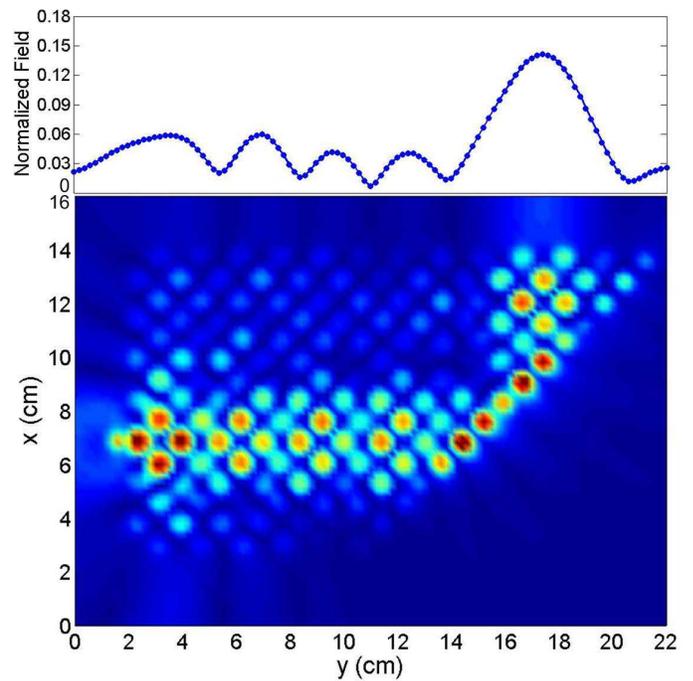}}
\caption{(Color online) Experimental beam propagation in the photonic crystal of 7.99mm$\pm$ 0.02mm rods at the self-collimation frequency (7.64~GHz) reflected from the surface of 5.92mm$\pm$ 0.01mm rods. The measured GH is approximately 4 lattice spacings. Upper plot shows the output field taken one lattice period beyond the surface.}
\end{figure}

Recently, we employed finite-difference time-domain (FDTD) calculations to study numerically the internal
Goos-H\"anchen (GH) shift observed at the reflection of a self-collimated beam from the modified surface of a truncated two-dimensional photonic crystal~\cite{Matthews:PLA-3098:2008}. By assuming that the surface rods can be modified or made of a nonlinear material, we demonstrated numerically a method for tuning the shift of the internally reflected beam by engineering the surface of the photonic crystal. In this Letter, we study the problem of beam reflection experimentally for truncated and engineered surfaces of a microwave photonic crystal fabricated from alumina rods and operating at microwave frequencies. This relatively simple experimental setup allows us to demonstrate a proof of principle of our theoretical concept for controlling the output position of a reflected beam and tunability of the internal GH shift through a change in the rod diameter at the reflection surface. Our experimental data are in good agreement with the results of the FDTD numerical calculations carried out for the experimental parameters.

In our experiment, we use a microwave photonic crystal created by a lattice of high purity alumina ceramic rods with the diameter of r=7.99mm$\pm$0.02mm and the refractive index of n$\approx$2.4 at the self-collimation frequency. These rods are set in a two-dimensional square lattice with a lattice spacing of a=10.59mm; the rods are placed in a frame of expanded polystyrene foam with n$\approx$1.01. This set up demonstrates the self-collimation effect for the lowest order TE mode at the frequency of 7.64GHz equating to a wavelength of $\lambda$=39.24mm.

Firstly, we employ these experimental parameters and study numerically the properties of the photonic crystal
using the planewave expansion method, implemented as Bandsolve, to produce the equal-frequency contours; this data allow us to determine the specific conditions for realizing the beam self-collimation effect.  Next, the numerical simulation results are used to determine an optimal structure to ensure the frequency region of the self-collimation effect are within the constraints of our experimental system. In particular, the self-collimation is directed at 45-degrees to the lattice direction and as such the structure is rotated in the experiment.

Next, we study numerically the reflection of self-collimated beams from the photonic-crystal surface
similar to the analysis reported earlier in Ref.~\cite{Matthews:PLA-3098:2008} for other parameters. We employ the FDTD method implemented as Fullwave, the propagation of a point source, set adjacent to the 6$^{th}$ surface rod from the short side of the photonic crystal, through our square lattice photonic crystals used in the experiments. In our study, the photonic crystal is 26 rods long by 15 rods wide with a cleaved surface at 45-degrees along the lattice direction. The cleaved surface layer is modified through a change in the diameter of the rods. The sets of rods tested experimentally have respective diameters of: 2.005mm$\pm$ 0.015mm, 2.98mm $\pm$ 0.01mm, 4.18mm $\pm$ 0.02mm, 4.915mm $\pm$ 0.005mm, 5.92mm$\pm$ 0.01mm, 7.175mm$\pm$ 0.015mm, $7.99$mm$\pm 0.02$mm. To measure the GH shift we take an output plane perpendicular to the input surface set at a distance of one lattice constant from the output surface. An example of the reflected beam experiencing the GH shift, calculated numerically, is shown in Fig. 1. For convenience we normalize the output peak magnitudes to emphasize the GH shift. The position of the input plane is translated to the center of the central reflection rod on the cleaved surface.

Figure 2 shows our experimental setup. The photonic crystal, consisting of high purity alumina ceramic rods, is held in place by a frame of expanded polystyrene foam, manufactured using a computer controlled 3D Milling Machine (Roland DG MDX-40). The completed photonic crystal is placed in a parallel plate waveguide (see Fig.~2) with a separation between the plates of $\sim$11mm with a source antenna of 1.3mm diameter and 10mm height in the lower plate, and an detection antenna set flush with the upper plate. A shaped carbon impregnated foam surrounds the measurement region to suppress reflections in the measurements. The lower plate (Fig.~2), including the input antenna and the photonic crystal, is mounted on a planar translation stage (Parker, 803-0936A, two-axis translation stage) to allow scanning of the field on the top surface of the photonic crystals. A vector network analyzer (Rohde and Schwarz model ZVB20) is used to provide both the input point source and receiving antenna to measure both the amplitude and phase of the transmission response (S12 parameter). This response is highly correlated to the vertical component of the electric field, equivalent to a TE mode.

\begin{figure}[htb]
\centerline{\includegraphics[width=10.5cm]{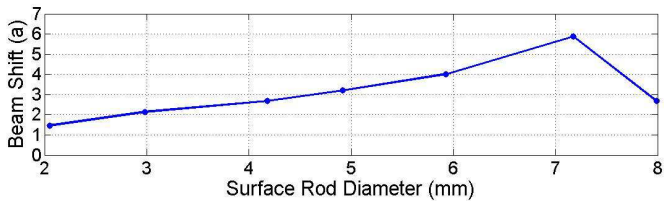}}
\caption{(Color online) The experimental GH shift (measured in lattice periods) for a set of different surface rods. The maximal shift occurs for a small reduction in the diameter of the surface rod (7.175mm$\pm$ 0.015mm) and the smallest shift occurs for a large reduction in the surface rod diameter (2.005mm$\pm$ 0.015mm).}
\end{figure}

Using this experimental setup, we study the internal beam reflection and measure the GH shift for different surface rods, reproducing well the results obtained by numerical FDTD simulations (cf. Fig.~3 and Fig.~1). In experiment,
the limitations of the antenna resolution lead to a smoothing of the double peak observed in the numerical simulations (Fig.~1) due to the averaging of the field from the surrounding area. The center position of the experimental peak is used to determine the GH shift for a given surface diameter.
The resulting peak shifts, measured in lattice constants relative to the perfect reflection center, are plotted in Fig.~4 for different values of the surface rods. We observe that the output beam position is sensitive to the diameter of the surface rods, as discussed earlier~\cite{Matthews:PLA-3098:2008}. In particular, we note that as the diameter of 7.99mm in our graph is the same as adding a layer with a surface diameter of 0mm. As such the shift at a rod diameter of zero should be 1 lattice constant less than the value at 7.99mm and from our results at 2.005mm it can be seen that this result is well reproduced.

The change of the effective index as a result of the change of the surface rod diameter provides us with an analogue to the nonlinearity-induced GH beam shift discussed earlier in our theoretical paper~\cite{Matthews:PLA-3098:2008}. While the effective index change used here to obtain a significant shift is large, a well-tuned system near a resonance could show a significant shift for a far smaller index change.

In conclusion, we have studied experimentally the Goos-H\"anchen shift experienced by
a self-collimated beam reflected from the surface of a truncated two-dimensional photonic crystal operating at microwave frequencies. We have demonstrated efficient control over the output position of the beam reflected through 90-degrees by modifying the surface rods and measuring the internal Goos-H\"anchen shift between one and six lattice constants. The experimental results are in a good agreement with the finite-difference time-domain numerical calculations.

This work was supported by an award under the Merit Allocation Scheme on the National Facility of the Australian Partnership for Advanced Computing and also by the Australian Research Council through the Center of Excellence Program. The authors thank David Powell and Ilya Shadrivov for useful discussions and help with microwave experiments.

\end{document}